\documentclass[arial,11pt]{article}
\usepackage{enumitem}
\usepackage{setspace}
\usepackage[a4paper,margin=1in]{geometry}
\usepackage{graphicx}
\usepackage{float}
\usepackage{footnote}
\usepackage{titlesec}
\usepackage{caption}
\usepackage{authblk}
\usepackage[super,comma]{natbib}
\makeatletter 
\renewcommand\@biblabel[1]{#1.} 
\makeatother

\titleformat{\section}{\normalfont\Large\bf}{}{0pt}{}
\titleformat{\subsection}{\normalfont\it}{}{0pt}{}
\begin{document}

\title{Data Adaptive Regularization for Abdominal Quantitative Susceptibility Mapping}
\author[1]{Julia V. Velikina} 
\author[1,2]{Ruiyang Zhao} 
\author[1,2]{Collin J. Buelo} 
\author[1]{Alexey A. Samsonov} 
\author[1,2,3,4,5]{Scott B. Reeder}
\author[1,2]{Diego Hernando}
\affil[1]{Department of Radiology, University of Wisconsin, Madison, WI, 53705, USA}
\affil[2]{Department of Medical Physics, University of Wisconsin, Madison, WI, 53705, USA}
\affil[3]{Department of Biomedical Engineering, University of Wisconsin, Madison, WI, 53705}
\affil[4]{Department of Medicine, University of Wisconsin, Madison, WI, 53705, USA}
\affil[5]{Department of Emergency Medicine, University of Wisconsin, Madison, WI, 53705, USA}
\date{}

\maketitle

\begin{center}

{\bf Correspondence to:}\\
Julia Velikina \\
Department of Radiology \\
University of Wisconsin-Madison \\
WIMR, 1111 Highland Dr, Rm. 1117 \\
Madison, WI 53705\\
USA \\
Phone:		1-608-265-2104 \\
Fax:		1-608-265-9840 \\
Email:		velikina@wisc.edu

\end{center}

\pagebreak 

\begin{abstract}
\noindent {\bf Purpose:} To improve repeatability and reproducibility across acquisition parameters and reduce bias in quantitative susceptibility mapping (QSM) of the liver, through development of an optimized regularized reconstruction algorithm for abdominal QSM.\\
{\bf Theory and Methods:} An optimized approach to estimation of magnetic susceptibility distribution is formulated as a constrained reconstruction problem that incorporates estimates of the input data reliability and anatomical priors available from chemical shift-encoded imaging. The proposed data-adaptive method was evaluated with respect to bias, repeatability, and reproducibility in a patient population with a wide range of liver iron concentration (LIC). The proposed method was compared to the state-of-the-art approach in liver QSM for two multi-echo SGRE protocols with different acquisition parameters at 3T. Linear regression was used for evaluation of QSM methods against a reference FDA-approved $R_2$-based LIC measure and $R_2^*$ measurements; repeatability/reproducibility were assessed by Bland-Altman analysis.  \\
{\bf Results:} The data-adaptive method produced susceptibility maps with higher subjective quality due to reduced shading artifacts. For both acquisition protocols, higher linear correlation with both $R_2$ and $R_2^*$-based measurements were observed for the data-adaptive method ($r^2=0.74/0.72$ for $R_2$, $0.98/0.99$ for $R_2^*$) than the standard method ($r^2=0.62/0.67$ and $0.84/0.91$). For both protocols, the data-adaptive method enabled better test-retest repeatability (repeatability coefficients 0.14/0.14ppm for the data-adaptive method, 0.26/0.31ppm for the standard method) and reproducibility across protocols (reproducibility coefficient 0.25ppm vs 0.36ppm) than the standard method. \\
{\bf Conclusions:} The proposed data-adaptive QSM algorithm may enable quantification of liver iron concentration with improved repeatability/reproducibility across different acquisition parameters as 3T.
\end{abstract}

\pagebreak

\section*{Introduction}
Excessive iron accumulation in the liver can lead to liver disease and eventual liver cirrhosis, hepatocellular carcinoma, diabetes mellitus or other endocrine disorders. Quantification of liver iron concentration (LIC) with low bias and variability is needed for the management of liver iron overload\cite{BrittenhamBadman2003}.  Liver biopsy is the most direct quantitative method of evaluating iron content; however, biopsy is an invasive procedure that carries its own risks, has limited reproducibility, and is not appropriate for long-term observations\cite{RatzChH}. Recently, quantitative susceptibility mapping (QSM) has emerged as a promising non-invasive technique for assessment of iron content in the liver\cite{Chu2004, HernandoSusc2013, Sharma2015}.

Magnetic susceptibility is a fundamental property of all materials, with iron being the major paramagnetic non-trace element that can detectably alter susceptibility in the body. The differences between magnetic susceptibilities of tissues lead to perturbation of the main magnetic field in MRI, which is encoded in the phase of complex acquisitions of gradient recalled echo (GRE) images. This relationship can be modeled as a convolution of susceptibility sources with the magnetic dipole kernel\cite{Kee2017}. QSM aims to solve the inverse problem of deriving local susceptibility distribution from the measured magnetic field perturbation. This inverse problem is ill-posed due to information loss in the forward problem, which results from vanishing of the Fourier transform of the dipole kernel along the double cone surface as well as the lack of information from areas outside the FOV (commonly conceptualized as the “background” field in QSM literature). A number of regularization techniques have been proposed to overcome the ill-posedness of QSM\cite{LiuMEDI2012, Langkammer2015, Rochefort2010}. 

Historically, QSM first demonstrated significant promise for characterization of paramagnetic ion deposition in brain tissues\cite{Wharton, Schweser2011, LiuMEDI2011}.   The success of QSM of the brain has been due in part to the ability to obtain a high resolution magnetic field map in moderate acquisition time and relative simplicity of the anatomy permitting to design reliable dipole kernel deconvolution methods. QSM in the abdomen, however, faces additional challenges, including more complex anatomy, presence of chemically-shifted fat, rapid signal decay (especially in cases of severe iron overload), and physiological motion. Specialized techniques for body QSM have been proposed\cite{Chu2004, HernandoSusc2013, Sharma2015, Sharma2017, WangHaselgrove1999} to address these challenges. Nevertheless, the presence of liver iron overload may lead to errors in the measured field map and complicate QSM preprocessing (e.g., the background field removal) and dipole inversion, potentially creating errors and amplifying artifacts such as shading in the resulting susceptibility maps. These artifacts often lead to bias as well as poor repeatability and reproducibility of QSM measurements. Additionally, limited scan time in breath-held acquisitions limits the achievable spatial resolution, which has been shown to introduce a bias into susceptibility measurements\cite{ZhouLiu2014, Karsa2019, ZhouColgan2019}. Therefore, there is an unmet need to improve the QSM algorithms in the body to enable low bias, high repeatability and high reproducibility.

The purpose of this work is to optimize the reconstruction of QSM of the abdomen, with a focus on quantification of liver iron overload. We formulate a constrained reconstruction problem that incorporates estimates of the input data reliability and anatomical priors available from chemical shift-encoded (CSE) imaging.  We then evaluate the new method with respect to bias, repeatability, and reproducibility across acquisition protocols. We perform this study in a patient population with a wide range of LIC and compare it to the state-of-the-art approach in liver QSM\cite{Sharma2015, Sharma2017}, which has been most thoroughly validated to date.  

\section*{Theory}
\subsection*{Problem Formulation and Regularized Solutions}
QSM estimates the magnetic susceptibility distribution from the measured main magnetic field, which itself is typically derived from phase accumulation in multi-echo gradient echo acquisitions. In the presence of water and fat signals, the multi-echo signal $s(r,t_n)$  acquired at echo times $t_n, n=1\ldots,N_{TE}$, can be modeled as follows:
\begin{equation}
s(r,t_n)=\left(s_{water} (r)+s_{fat} (r) \sum_{j=1}^6 \rho_j e^{2\pi i\Delta f_j t_n} \right) e^{-R_2^* (r) t_n } e^{2\pi i \psi(r) t_n},
\label{FW_signal_model}
\end{equation}
where $r$ is pixel location, $s_{water}$ and $s_{fat}$ are water and fat signals, respectively, $\Delta f_j$ and $\rho_j$ are frequency shifts of fat species relative to water and their relative amplitudes\cite{YuReeder2008}, respectively, $R_2^*$ is transverse relaxation rate, and $\psi$ is the static field inhomogeneity. This process enables estimation not only of the field map but also fat and water images, $R_2^*$, and fat fraction maps. Based on model fitting of this signal model, the residual of the fit can be described as follows:
\begin{equation}
{\cal E}(s;r) = \left( \sum_{n=1}^{N_{TE}} \left|s(r,t_n) - \left(s_{water} (r)+ s_{fat}(r) \sum_{j=1}^6 \rho_j e^{2\pi i\Delta f_j t_n} \right) e^{-R_2^* (r) t_n } \right|^2 \right)^{1/2}.
\label{FW_fit_residual}
\end{equation}
If the problem is over-determined, the residual provides a measure of reliability of the parametric estimates, including the local field map estimates.

Local susceptibility-induced variations in the main magnetic field can be modeled as a convolution (denoted by the $\ast$ symbol) of the susceptibility distribution $\chi$ with the dipole kernel\cite{WangLiu2015} 
\begin{equation}
\psi_{loc}=\frac{\gamma}{2\pi} B_0  d\ast\chi,
\label{Local_susc}
\end{equation}
where $\gamma/2\pi$ is the gyromagnetic ratio, $B_0$ is the strength of the main magnetic field, and the dipole kernel is given by $d(r)=\frac{3\cos^2\theta -1}{4\pi |r|^3}$, where $\theta$ denotes the angle of the position vector $r$ with respect to the direction of the main magnetic field. Determination of $\chi$ from Eq.\,(\ref{Local_susc}) via a direct deconvolution in $k$-space is ill-posed as the Fourier transform of the dipole kernel vanishes along the surface of a cone $\Gamma_0$   in $k$-space defined by $\frac{k_z^2}{|k|^2}=\frac{1}{3}$, causing extreme sensitivity to numerical and data errors. Rather, the susceptibility distribution can be obtained as a regularized least squares solution of Eq.\,(\ref{Local_susc}):
\begin{equation}
\chi = \arg \min_\chi  \left( \| (\psi_{loc}-D\ast\chi)\|_2^2  + \lambda \| P\chi \|_2^2 \right),
\label{Reg_min}
\end{equation}
where $D(r)=\frac{\gamma}{2\pi} B_0  d(r)$, $P$ is a regularization operator (typically, gradient-based) and $\lambda$ is a regularization parameter. However, lack of information along $\Gamma_0$ leads to poor conditioning of the inverse problem and typically results in artifacts manifesting as streaking and shading\cite{Kee2017}.

\subsection*{Background Field Removal and Data Weighting}

Determination of susceptibility distribution is further complicated by the fact that the measured field map comprises both local and background susceptibility-induced fields as well as shim fields and other sources of inhomogeneity. The desired local component can be explicitly estimated from the total measured field map prior to dipole inversion by a method such as SHARP\cite{Schweser2011}, LBV\cite{ZhouLiuLBV2014}, PDF\cite{LiuKhalidovPDF2011, Rochefort2010}, and others\cite{HorngSharma2016}. Each of these approaches has its own limitations described in a review by Schweser et al\cite{SchweserReview2017}.  Alternatively, extraction of the local field can be done implicitly relying on the assumption that the contributions to the total field map from the background susceptibility-induced field and other sources are described by harmonic functions\cite{LiLeigh2001}, which are annihilated by the Laplace operator $L$, hence $L\psi = L\psi_{loc}$.  

The accuracy of the measured field map may be further compromised by estimation errors, due to air (e.g., in the lungs), physiological motion, including respiration, cardiac motion, or peristalsis, and rapid signal decay due to high iron concentration in the liver and other organs.  The ill-conditioning of the inverse problem results in amplification of these errors, which can be counteracted by application of a weighting term W reflecting the reliability of the field map estimate at each voxel. By including this weighting term $W$, Eq.~(\ref{Reg_min}) is rewritten for simultaneous background field removal and susceptibility distribution estimation\cite{Sharma2015} as:
\begin{equation}
\chi = \arg \min_\chi  \left( \| WL (\psi_{loc}-D\ast\chi)\|_2^2  + \lambda \| P\chi \|_2^2 \right).
\label{Reg_minW}
\end{equation}

In the current state-of-the-art QSM\cite{Sharma2015}, $W$ is selected to be
\begin{equation}
W = \left(\sum_{n=1}^{N_{TE}} |s(r,t_n)|^2 \right)^{1/2}
\label{Wsos}
\end{equation}
to compensate for non-uniform noise variance in the field map estimates (as the noise standard deviation in the field map estimate is assumed to be inversely proportional to the SNR of the magnitude images\cite{RicianMRInoise}).    However, such choice of $W$ is oblivious to areas of relatively high signal with field map estimation errors due to motion artifacts or other sources. Here, we propose to leverage the fact that in CSE imaging the number of acquired complex-valued images is typically greater than the number of parameters to be estimated (e.g., a multi-echo acquisition of six echoes results in 12 real/imaginary data points used to estimate six real-valued paramerers: fat and water phase and amplitude, $R_2^*$, and field map). The overdetermined nature of the field map estimation problem allows  using residuals of the multi-echo CSE modeling as given by Eq.~(\ref{FW_fit_residual}) to detect estimation errors. We propose to modulate the weighting matrix by the size of the residual error as follows: 
\begin{equation}
W = \left(\sum_{n=1}^{N_{TE}} |s(r,t_n)|^2 \right)^{1/2} / {\cal E}(s;r).
\label{Wresidual}
\end{equation}	
This formulation aims to account for field map estimation uncertainties and to counteract amplification and propagation of these errors in the iterative susceptibility estimation.

\subsection*{Automated Selection of Reference Tissue}

The formulation of Eq.~(\ref{Reg_minW}) with the regularization operator $P$ based on the image gradient as in\cite{Sharma2015, LiuMEDI2011} produces a solution for $\chi$ only up to an additive constant. Therefore, susceptibility quantification is usually performed relative to some reference tissue, for instance, CSF in brain QSM\cite{LiuSpincemailleMEDI0}. This is performed by placing ROIs in the tissue of interest and in the reference tissue and subtracting the QSM measurements. In liver imaging, adipose tissue is typically chosen as a reference tissue as it is ubiquitous in the abdomen, does not accumulate iron, and has uniform susceptibility\cite{KnutsonWessling2003}. While the use of fat as a reference is a popular choice in most previous liver QSM methods\cite{Sharma2015}, it has its own shortcomings: although an ROI is usually placed in adipose tissue next to the liver ROI, the presence of residual shading in the susceptibility map may affect the measurements and lead to errors and poor repeatability/reproducibility. Therefore, in this work we propose to incorporate constraints on the susceptibility of adipose tissue directly into the dipole inversion in order to (i) avoid the need for reference ROI measurements; (ii) suppress the shading and measurement errors associated with poor susceptibility estimates in fat. In particular, we propose to incorporate susceptibility priors $\chi_0$ in spatial locations defined by a mask $M$ as an additional regularization term: 
\begin{equation}
\chi = \arg \min_\chi  \left( \| WL (\psi_{loc}-D\ast\chi)\|_2^2  + \lambda \| P\chi \|_2^2 + \mu \|M(\chi-\chi_0)\|_2^2 \right).
\label{Reg_minW_mask}
\end{equation}
The availability of a fat fraction (FF) map that can be determined with sufficient confidence from the CSE fat/water separation allows for determination of the fat mask $M=M_{fat}$ that is perfectly co-localized with susceptibility maps and does not require additional acquisition or processing time.  Details of the determination of the fat mask are described in Methods below.	
\section*{Methods}
\subsection*{Data Acquisition}

After IRB approval and obtaining informed written consent, 56 human subjects with known or suspected iron overload were scanned on a 3.0T clinical MRI system (MR750 or Premier, GE Healthcare, Waukesha, WI, USA). For each subject, two different multi-echo 3D SGRE acquisitions were performed during a breath hold, in order to evaluate reproducibility across relevant acquisition parameters. These two protocols (see details in Table~\ref{AcqParams}) will be referred to as High Resolution/Longer TE (HR/LTE) and Low Resolution/Shorter TE (LR/STE), respectively. In 35 of these exams, for evaluation of repeatability, the subject was removed from the scanner, the anterior coil array removed, the subject was asked to sit up and lie back down, the coil was replaced, localizer acquisitions were repeated, and the same multi-echo SGRE acquisitions were repeated. Finally, each subject was scanned using a 2D multi-slice spin-echo sequence at 1.5T. The spin-echo data were acquired to perform $R_2$-based liver iron quantification (Ferriscan, Resonance Health, Perth, Australia)\cite{StPierre2004}.

\begin{table}[h!]
   \begin{center}
	\caption{Acquisition parameters of the imaging protocols performed in the study.${}^a$}
	\label{AcqParams}
	\scriptsize
	\begin{tabular}{|p{0.48in}|p{0.52in}|p{0.31in}|p{0.3in}|p{0.6in}|p{0.2in}|p{0.35in}|p{0.7in}|p{0.65in}|p{0.21in}|p{0.33in}|}
	  \hline
	   \bf{Name} & \bf{Pulse}       & \bf{Scan} & $\mathbf{N_{TE}/}$ & ${\rm\bf TE}_1/\mathbf{\Delta} {\rm \bf TE}$ & \bf{TR} & \bf{FOV}            & \bf{Matrix} & \bf{Voxel}           & \bf{FA }  & \bf{RBW} \\
	        { }       & \bf{sequence} & {\bf time}   & \bf{ETL}  & (ms)                                                           & (ms)      & (${\rm cm}^2$) & \bf{size}     & \bf{size}  (${\rm mm}^2$)            & (${}^\circ$)  & \bf{ (kHz)}              \\
            \hline
	  Pr. 1, HR/LTE & 3D ME SPGR, 3T & 20 s& 6/3 & 1.2/1.0 & 8.0 & 40 x 32 & 256 x 144 x 32 & 1.6 x 2.2 x 8 & 3 & $\pm 125$ \\
	  \hline
	  Pr. 2, LR/STE & 3D ME SPGR, 3T & 19 s& 8/4 & 0.65/0.6 & 6.0 & 40 x 32 & 144 x 128 x 32 & 2.8 x 2.5 x 8 & 9 & $\pm 125$ \\
	  \hline
	  R2-based LIC & 2D spin-echo, 1.5T & 1020 s& 5/1 & $6/3$ & 1000 & 44 x 33 & 256 x 192 x 11 & 1.7 x 1.7 x 6 & 90 & $\pm 62.5$ \\
	  \hline
	\end{tabular}
          \vspace{11pt}\\
           \footnotesize{${}^a \mathbf{N_{TE}}=$ number of echoes; {\bf ETL} = echo train length; {\bf FA} = flip angle; {\bf RBW} = receiver bandwidth.}
   \end{center}
\end{table}

\subsection*{QSM Reconstruction}

Each multi-echo SGRE acquisition was processed using a CSE signal model (Eq.~(\ref{FW_signal_model}) above) to estimate the field map ($\psi$) as well as $R_2^*,  s_{water}$, and $s_{fat}$. A map of FF was estimated from $s_{water}$ and $s_{fat}$. In order to avoid fat-water swaps, the field map estimation was initialized using a regularized field map estimation method with a graph-cut algorithm\cite{HernandoGraphCut2010}.  

Susceptibility mapping was performed using two different methods: 
\begin{enumerate}[label=\arabic*)]
\item 	The state-of-the-art liver QSM method of Eq.~(\ref{Reg_minW}) described in\cite{Sharma2015}, which features $W$ defined in Eq.~(\ref{Wsos}) and a single regularizing operator $P$, based on image gradient. Since all regularization is based on information from source images only, we refer to this method as {\it spatially regularized}.
\item The proposed method defined by Eqs.~(\ref{Wresidual}-\ref{Reg_minW_mask}). As this method incorporates different sources of information available from CSE imaging in addition to spatial regularization, we refer to it as {\it data adaptive}. For the proposed method, the fat mask $M_{fat}$ comprised pixels within the body with ${\rm FF} > 0.9$ and $R_2^*<300\,{\rm s}^{-1}$. The latter condition helps avoid pixels with fat/water swaps that are possible in cases with iron overload (since adipose tissue does not accumulate iron, its $R_2^*$ values should not be elevated even in patients with  iron overload).  Additionally, to assess the choice of suitable reference tissue for abdominal QSM, we also performed mapping with the proposed data adaptive method but using $M=M_{aorta}$, based on the manually segmented abdominal aorta\cite{Jafari2019} and compared both to QSM without zero-referencing (choosing the mask to be an empty set, $M=\emptyset$). 
\end{enumerate}
Both methods were implemented in MATLAB 2018 (MathWorks, Inc., Natick, Massachusetts, USA) using a conjugate gradient algorithm. Iterations continued until either the maximum number of iterations ($N = 500$) was reached or the residual error relative to the data fell below a preset tolerance ($10^{-9}$). Regularization parameters in Eqs.~(\ref{Reg_minW},\ref{Reg_minW_mask}) were optimized empirically based on image sharpness and artifact reduction and then fixed in all experiments.
\subsection*{Data Measurements and Analysis}
	For each QSM method in each acquisition, susceptibility values in the liver were quantified by placing ROIs in a single slice of the right liver lobe in the segments VI or VII. The mean value of the ROI radii was 1.4~cm. In the method without embedded zero-reference regularization, quantification was performed relative to the average susceptibility in an ROI placed in the same slice in the nearby subcutaneous adipose tissue. Depending on the amount of adipose tissue in each subject, the radii of these ROIs ranged from 0.4~cm for very lean subjects to 1.7~cm,  with the mean radius value of 0.95~cm. For each reconstruction approach, linear regression analysis was performed to determine the correlations between liver susceptibility and liver $R_2^*$ values computed in the same liver ROIs as well as with $R_2$-based estimates of LIC. Repeatability and reproducibility of each QSM method was assessed using linear regression and Bland-Altman analysis\cite{BlandAltman} to report the repeatability coefficients (RC) defined as $RC = 1.96 \times \sqrt{2\sigma^2}$, where $\sigma$ is within-subject standard deviation; and reproducibility coefficients (RDC), defined analogously with $\sigma^2$ corresponding to within-subject variance under reproducibility conditions\cite{QIBA_terminology}. RC/RDC predict that the absolute difference between two measurements on a subject will differ by no more than the RC/RDC value on $95\%$ of occasions. The severity of shading artifacts in the susceptibility maps was evaluated by considering histograms of susceptibility values. The reduction of shading artifact corresponds to histograms with narrower, less overlapping modes\cite{TomazevicLikar2002}.

\section*{Results}
For six subjects, significant image artifacts prevented reliable measurements of susceptibility from the data acquired with Protocol 1 (HR/LTE). The nature of the artifacts was related to high iron contents of the liver, which led to large fat-water separation error and, consequently, to large errors in field map estimation (in one of these cases, $R_2$-based LIC quantification failed as well). The same problem was present in the data acquired with Protocol 2 (LR/STE) for three subjects only, likely due to the fact that shorter echo times led to less $R_2^*$  decay. These subjects were excluded from the analysis. In the remaining subjects, QSM reconstruction was successful. In several cases, where ghosting artifacts were present due to poor breath-hold, measurements were performed in the regions that avoided artifacts.

\begin{figure}[h!]
  \begin{center}
    \includegraphics{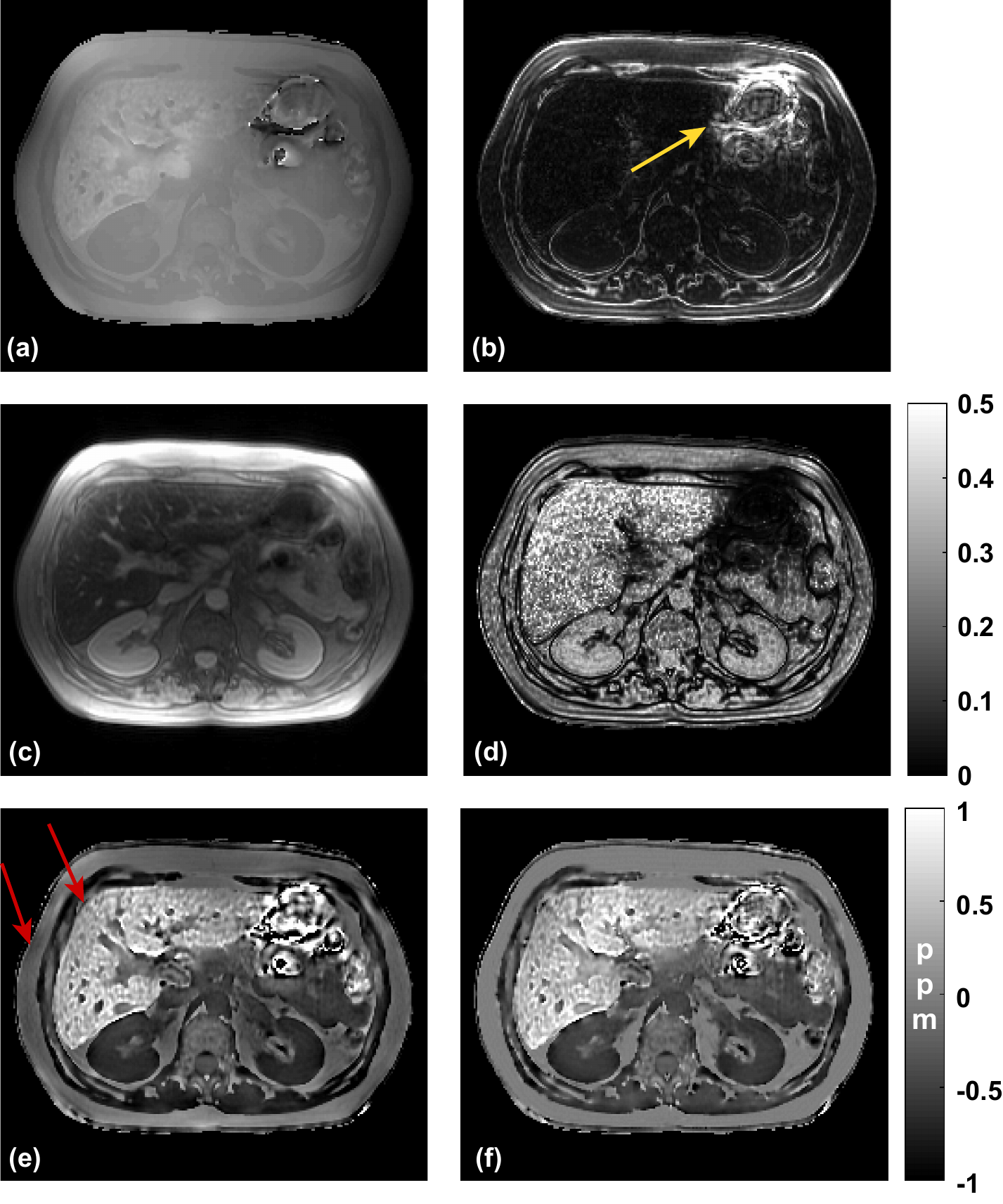}
  \end{center}

  \caption{(a) Field map for a single slice in a patient with elevated liver iron content. Note areas of poor field map estimation near the intestine, likely due to the presence of air and motion. (b) Residual of CSE fat/water separation fit. Note areas of poor fit near the intestine (arrow) as well as in partial volume voxels (tissue boundaries). (c) Sum of squares image used for data weighting in the spatially constrained method. (d) Residual-based data weighting of the data adaptive method. (e) Susceptibility map obtained with the spatially constrained method. Note shading in the liver and subcutaneous fat (arrows). (f) Susceptibility map from the data adaptive method exhibits improved signal homogeneity both in the liver and subcutaneous fat.}
  \label{Fig_map_examples}
\end{figure}

Images in Fig.~\ref{Fig_map_examples} provide a typical example of challenges encountered in abdominal QSM. These include errors in field map estimation, which are often due to large susceptibility differences at air/tissue interfaces, e.g. next to the intestine and lungs, and motion (Fig.~\ref{Fig_map_examples}a). These errors can be further confirmed by examining the residual of CSE fat/water separation fit (Fig.~\ref{Fig_map_examples}b). Using the standard weighting of the data fidelity term in Eq.~(\ref{Reg_minW}) as defined in Eq.~(\ref{Wsos}) (Fig.~\ref{Fig_map_examples}c) implies the similar level of confidence both in well-estimated regions of the field map in the liver and in the poorly estimated regions. In contrast, the proposed data adaptive weighting of Eq.~(\ref{Wresidual}) downplays the problematic regions (Fig.~\ref{Fig_map_examples}d), reflecting confidence in the field map estimate used as a QSM input. This has a direct effect on the estimated susceptibility maps (Fig.~\ref{Fig_map_examples}e-f). While the area of motion remains unrecovered, its impact on overall inversion is reduced in the proposed method as evidenced by improved homogeneity and reduced shading in the liver and subcutaneous fat, which affect the spatially constrained method (arrows).

\begin{figure}[h!]
  \begin{center}
    \includegraphics[width=\textwidth]{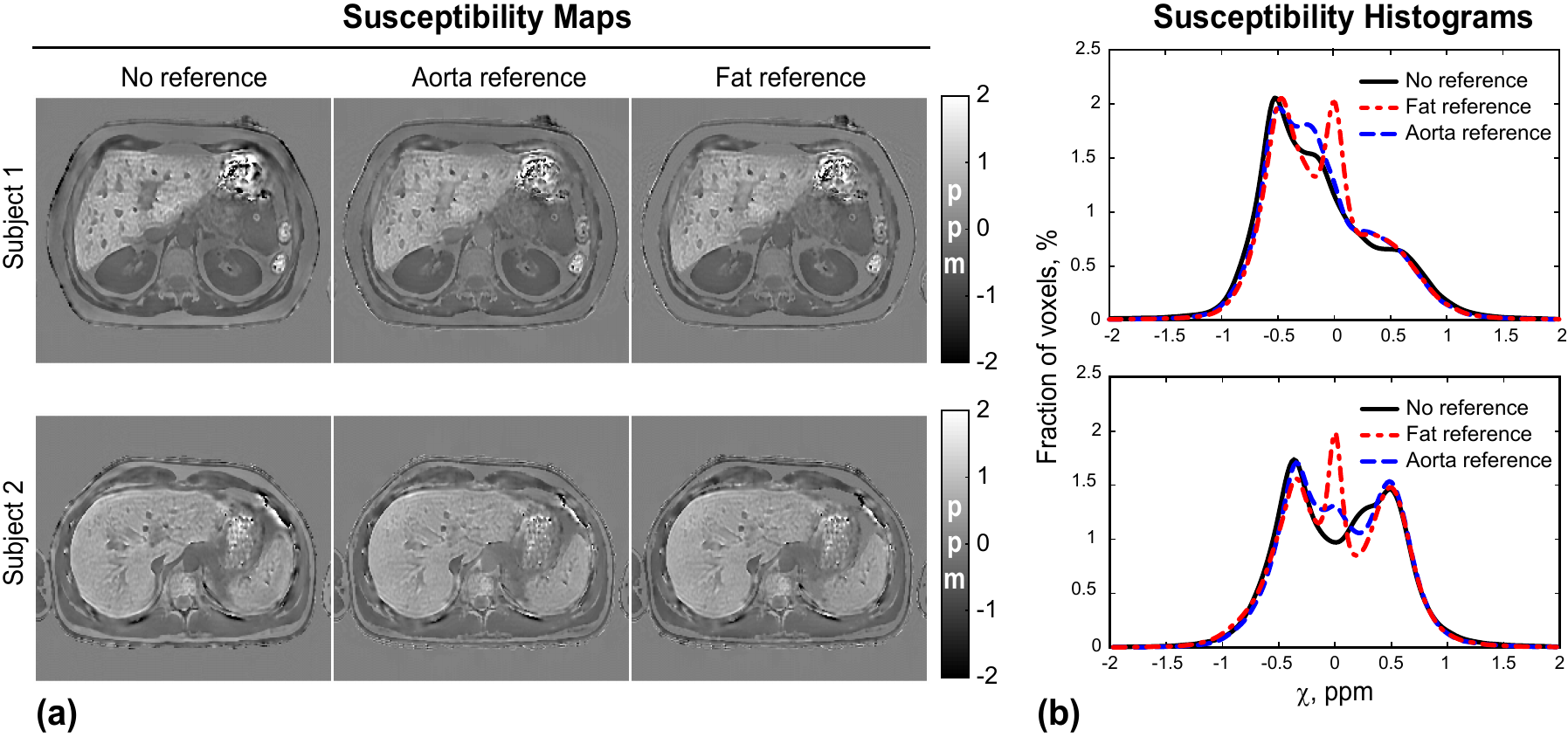}
  \end{center}

  \caption{(a) Comparison of susceptibility maps obtained in two human subjects with different amounts of subcutaneous adipose tissue without a reference tissue regularization term as well as using abdominal aorta and adipose tissue as zero-reference. The use of adipose tissue as zero-reference provides the most reduction in shading artifact even in the case of a lean subject (subject 2, bottom row) as evidenced by well-defined peaks of clearly separated modes in its histogram (b).}
  \label{Fig_histograms}
\end{figure}

Comparison of susceptibility maps obtained without zero-referencing ($M=\emptyset$ in Eq.~(\ref{Reg_minW_mask})) with those that use abdominal aorta or adipose tissue as zero-reference ($M=M_{aorta}$ or $M=M_{fat}$ in Eq.(\ref{Reg_minW_mask})) in Fig.~\ref{Fig_histograms}a demonstrate that the latter approach leads to most reduction of the shading artifact. This is further confirmed by histograms of the susceptibility values, which exhibit well-defined peaks (Fig.~\ref{Fig_histograms}b) when adipose tissue is used for zero-referencing. Clearly separated modes with less overlap indicate a reduction in shading in susceptibility maps.

\begin{figure}[h!]
  \begin{center}
    \includegraphics{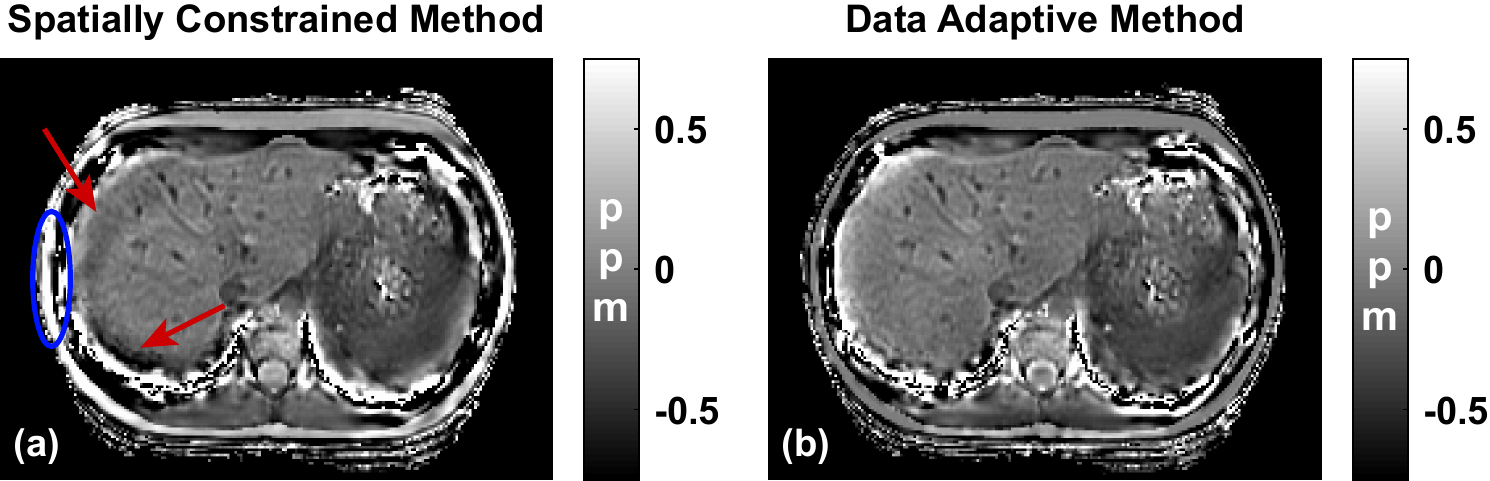}
  \end{center}

  \caption{Shading artifact in both liver (red arrows) and subcutaneous fat (blue oval) of the susceptibility map obtained with the spatially constrained method (a) that leads to unreliable liver susceptibility measurements is removed in the map obtained with the data adaptive method (b).}
 \label{Fig_susc_example}
\end{figure}

Figure~\ref{Fig_susc_example} provides an example of shading artifact in susceptibility maps obtained with the spatially constrained method (especially pronounced in the subcutaneous adipose tissue), which makes quantification of susceptibility dependent on ROI placement and, therefore, unreliable. This artifact is rectified in the data adaptive method.

\begin{figure}[h!]
  \begin{center}
    \includegraphics[width=\textwidth]{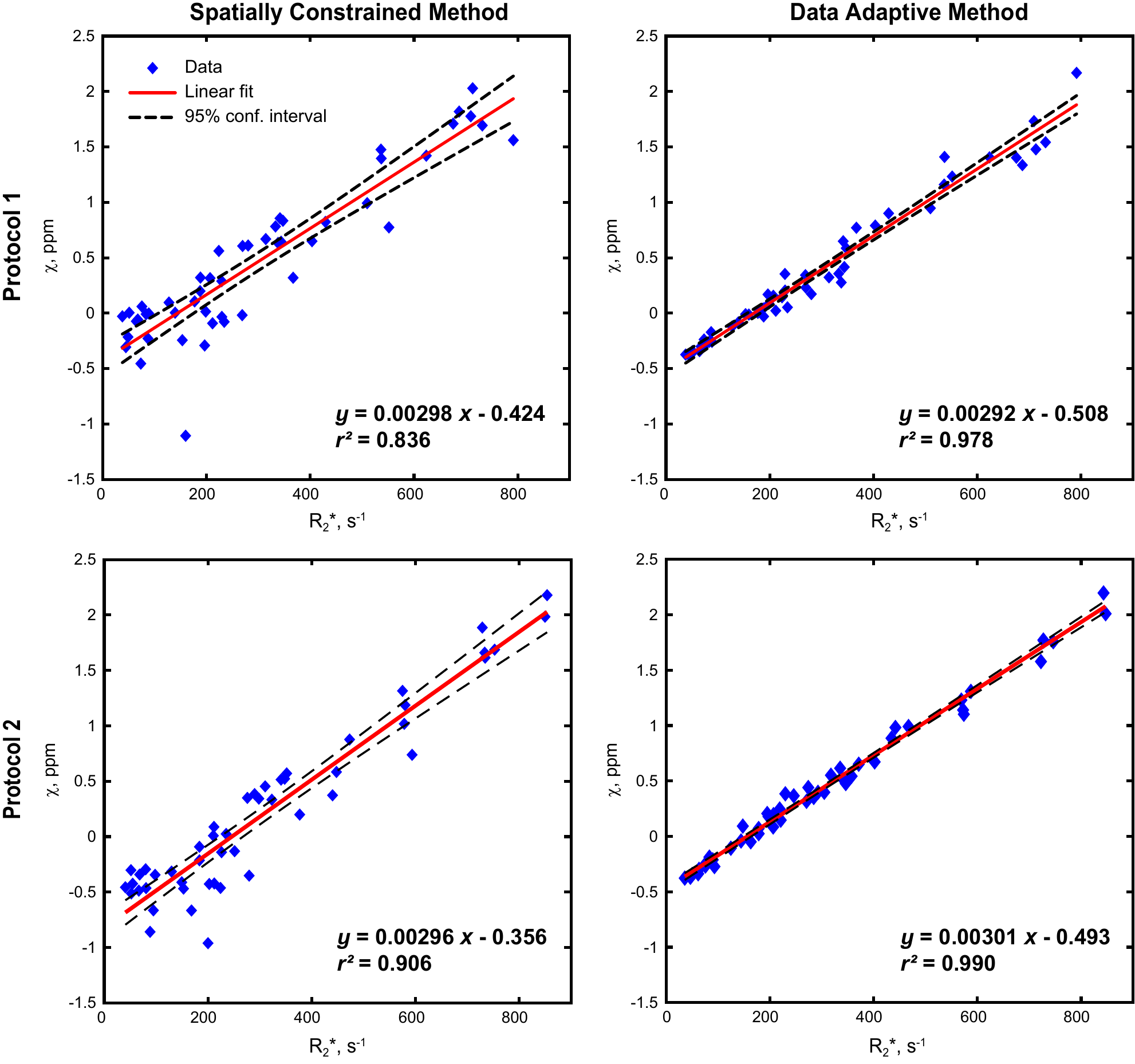}
  \end{center}

  \caption{Linear regression analysis indicates that, compared to the spatially constrained method, the data adaptive method has higher correlation of the liver susceptibility values with $R_2^*$ for both HR/LTE (top row) and LR/STE (bottom row) acquisition protocols}
 \label{Fig_R2st_corr}
\end{figure}

\begin{figure}[h!]
  \begin{center}
    \includegraphics[width=\textwidth]{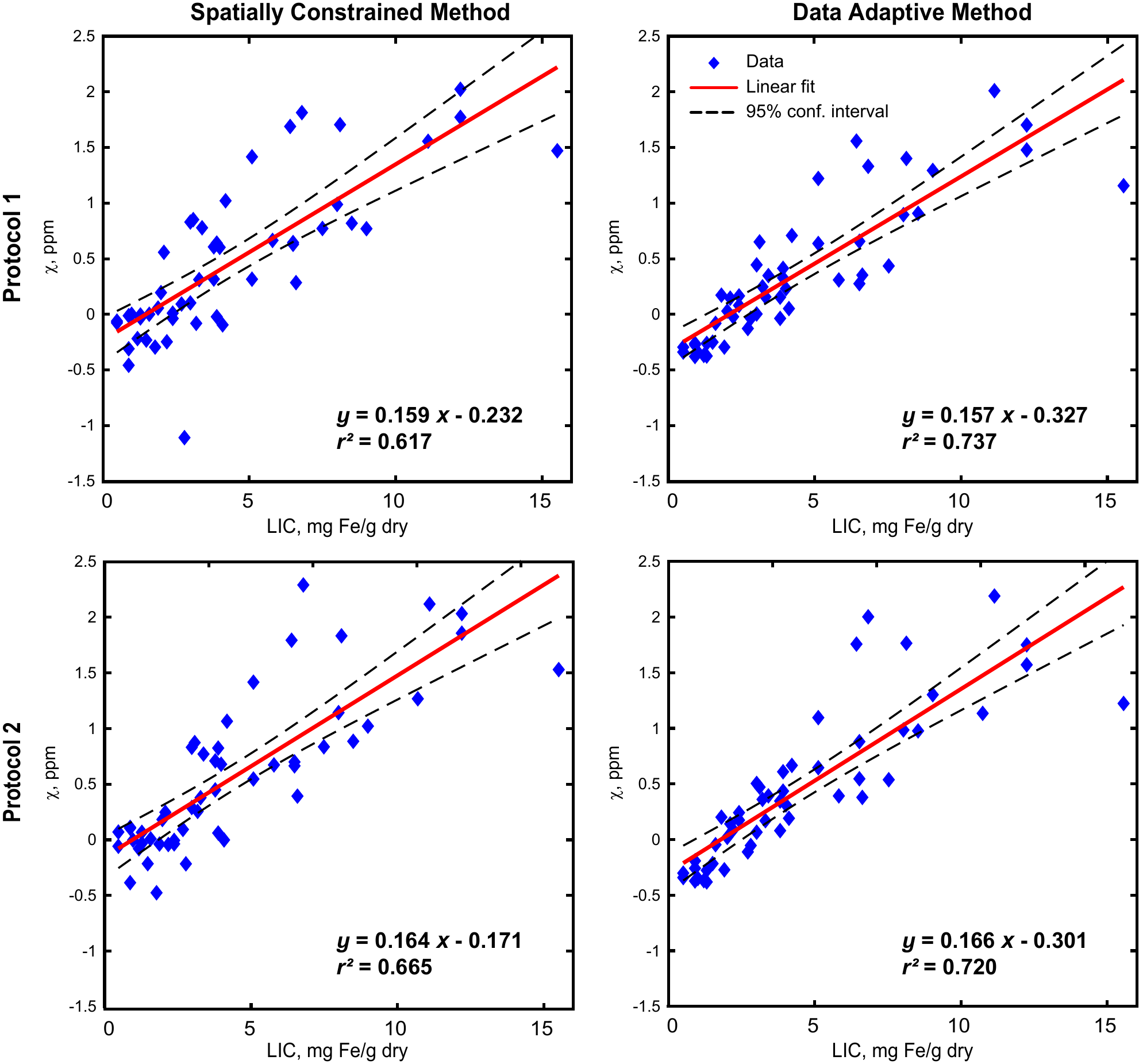}
  \end{center}

  \caption{Linear regression analysis of the relationship between $R_2$ relaxometry-based LIC and liver susceptibility values demonstrates higher correlations for the data adaptive method than the spatially constrained method for both HR/LTE (top row) and LR/STE (bottom row) acquisition protocols.}
 \label{Fig_ferriscan_corr}
\end{figure}

The linear regression analysis of the relationship between susceptibility values and $R_2^*$ measured in the liver for Protocol 1 (Fig.~\ref{Fig_R2st_corr}a-b) demonstrates higher correlation with susceptibility values obtained with the data adaptive method ($r^2= 0.98$) as compared to the spatially constrained ($r^2= 0.84$). Similarly, for Protocol 2 (Fig.~\ref{Fig_R2st_corr}c-d), the data adaptive method results in higher correlation ($r^2= 0.99$) between liver susceptibility values and $R_2^*$ than the spatially constrained method ($r^2= 0.91$). The same trend is observed (Fig.~\ref{Fig_ferriscan_corr}) in the linear regression analysis between susceptibility values in the liver and $R_2$-based LIC, with the data adaptive method exhibiting higher correlations for both acquisitions ($r^2= 0.74/0.72$) than the spatially constrained ($r^2 = 0.62/0.66$). Additionally, in  all cases the data adaptive method demonstrates tighter $95\%$ confidence intervals of the regression coefficients. 

\begin{figure}[h!]
  \begin{center}
    \includegraphics{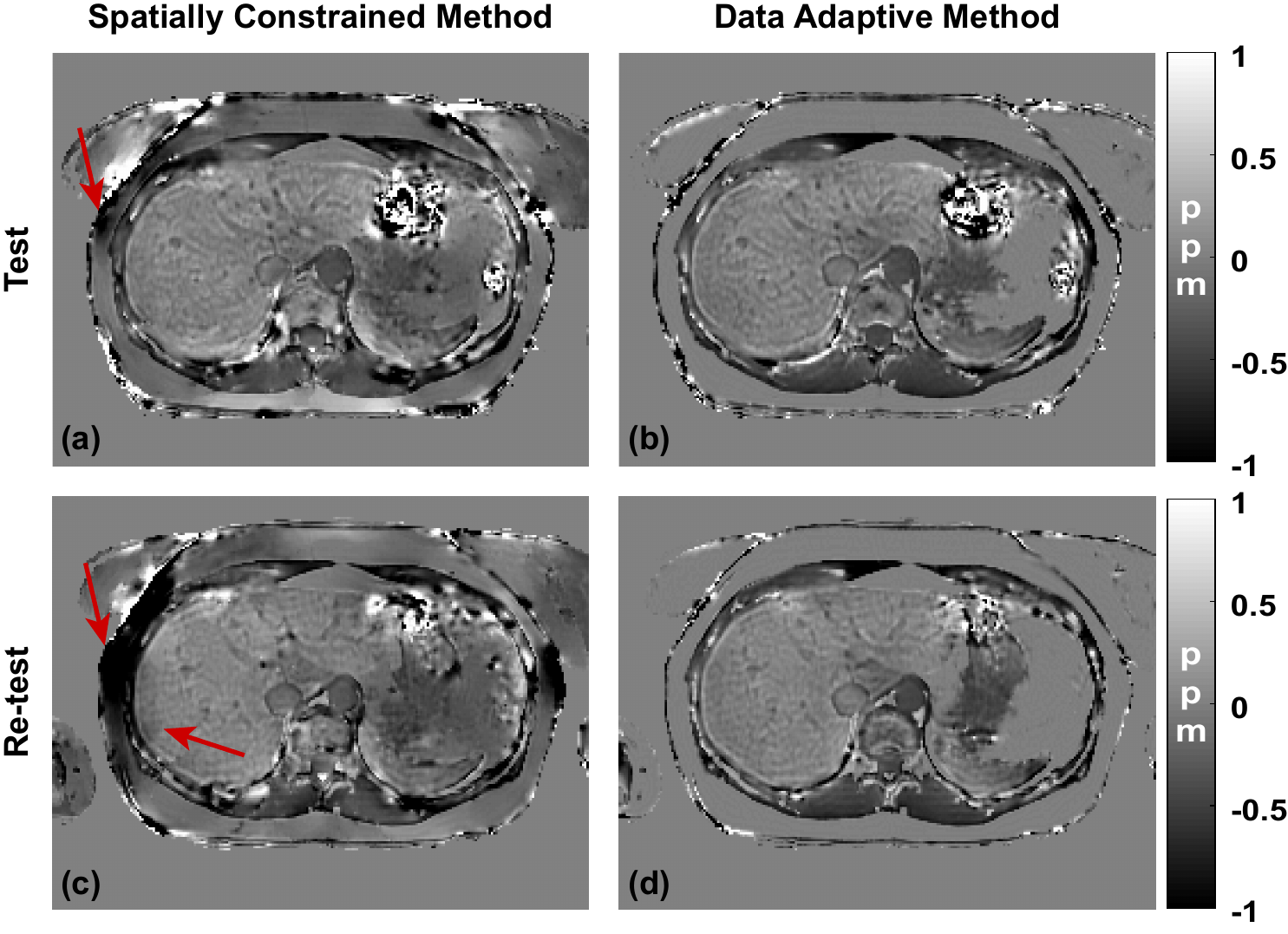}
  \end{center}

  \caption{Compared to the spatially constrained liver QSM method (a: test, c: re-test), the proposed data adaptive method (b: test, d: re-test) produces more consistent (repeatable) susceptibility maps.}
 \label{Fig_repeat_ex}
\end{figure}

\begin{figure}[h!]
  \begin{center}
    \includegraphics[width=\textwidth]{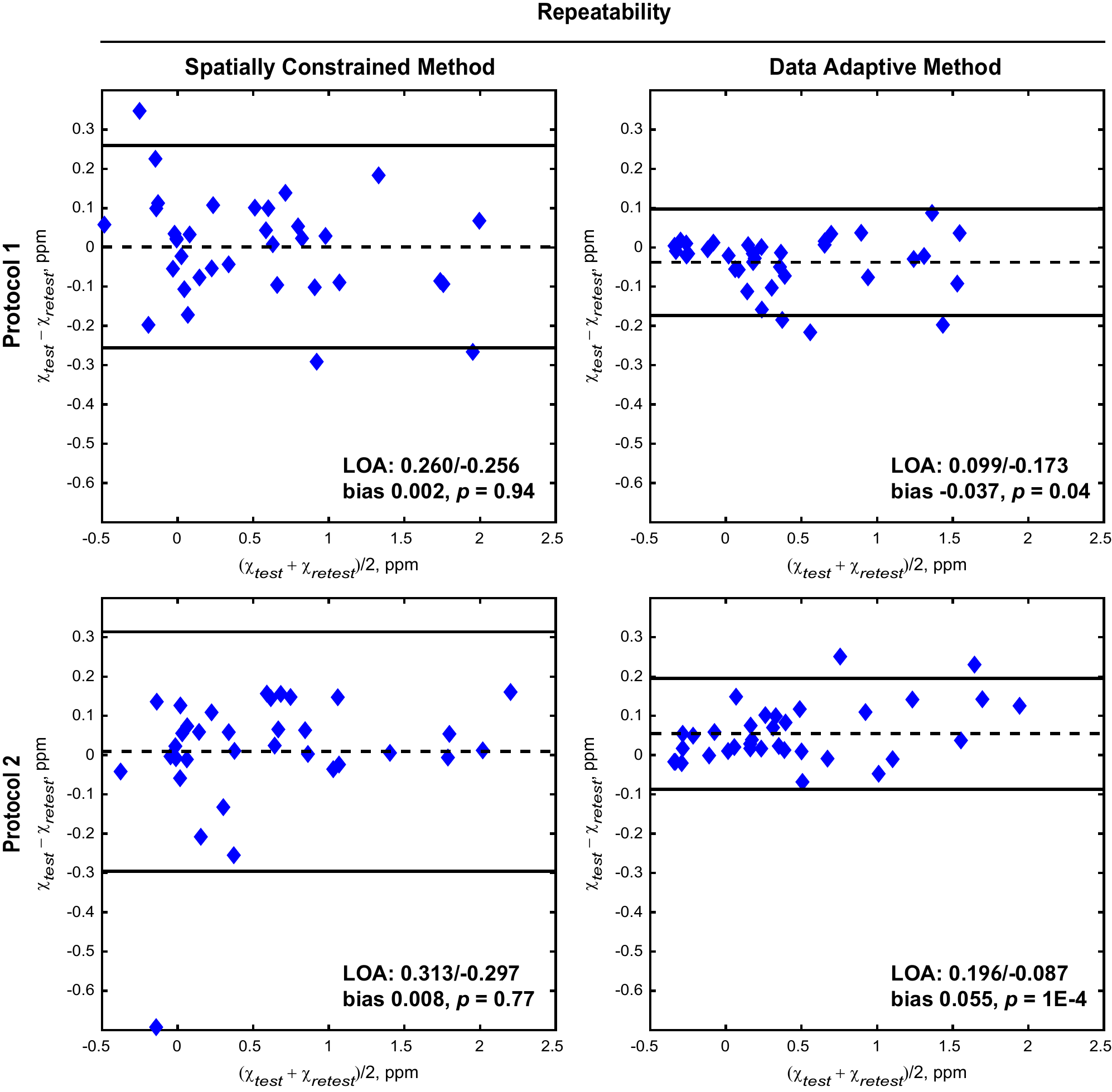}
  \end{center}

  \caption{Bland-Altman plots indicate higher repeatability of the data adaptive method for both protocols with tighter limits of agreement (LOA) and small statistically significant bias than the spatially constrained method.}
 \label{Fig_repeat_plots}
\end{figure}

\begin{figure}[h!]
  \begin{center}
    \includegraphics[width=\textwidth]{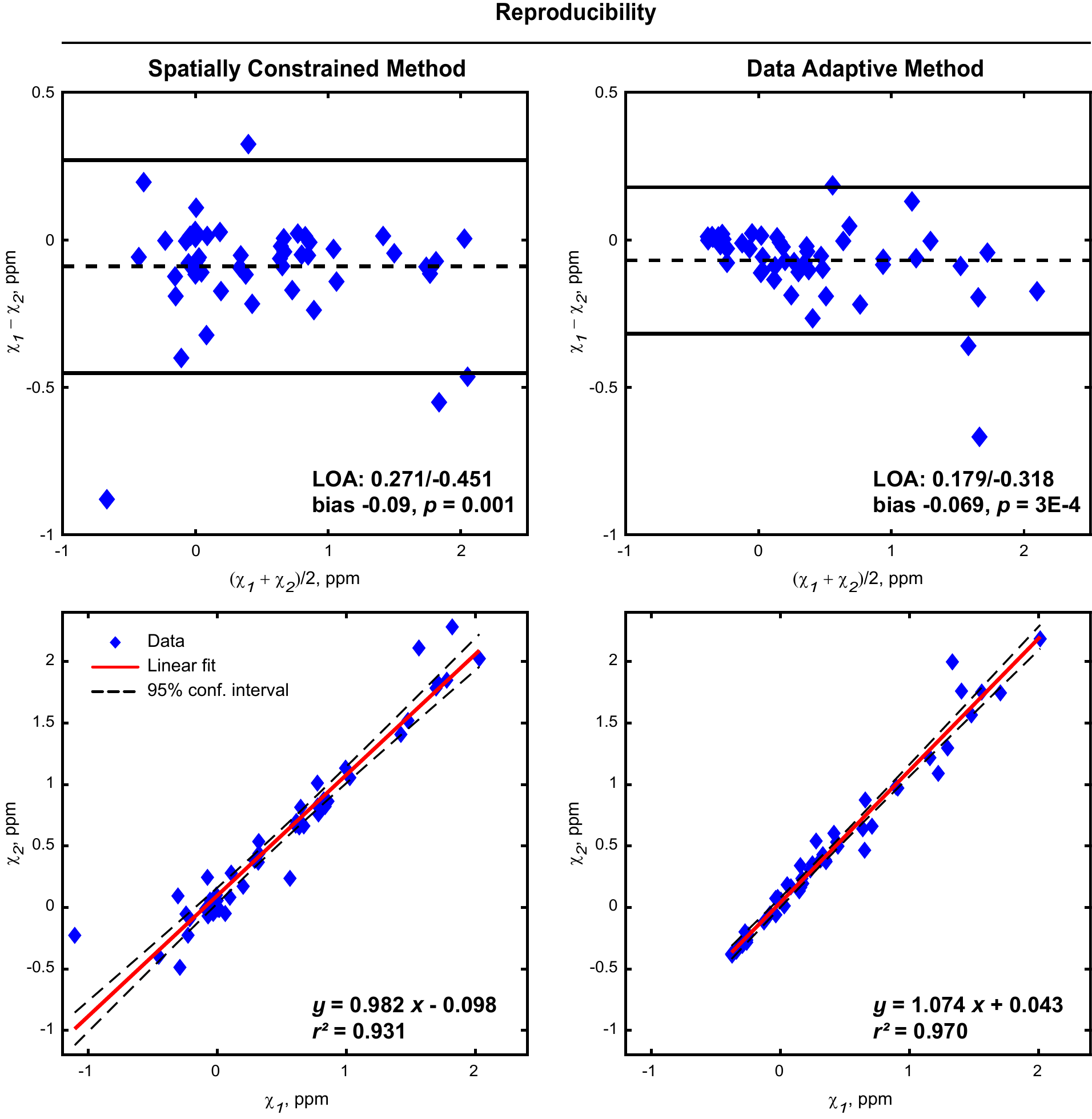}
  \end{center}

  \caption{Bland-Altman plots (top row) indicate higher reproducibility of the proposed data adaptive method (RDC=0.25 ppm) than the spatially constrained method (RDC=0.36 ppm) across two scanning protocols with different spatial resolution. This is further confirmed by linear regression analysis that shows a higher correlation of the susceptibility values obtained with the data adaptive method ($r^2=0.970$) than with the spatially constrained one ($r^2=0.931$).}
 \label{Fig_repr_plots}
\end{figure}

As illustrated in Fig.~\ref{Fig_repeat_ex}a,c, the shading artifact (arrows) in the spatially constrained method may lead to poor repeatability between consecutive test and re-test acquisitions, while the data adaptive method produces a more consistent reconstruction (Fig.~\ref{Fig_repeat_ex}b,d). As a result, the data adaptive method has substantially improved repeatability (RC = 0.14/0.14 ppm for protocols 1 and 2) compared to the spatially constrained method (RC = 0.26/0.31 ppm) (Fig.~\ref{Fig_repeat_plots}). Finally, as illustrated in Fig.~\ref{Fig_repr_plots}, the data adaptive method also shows higher reproducibility (RDC = 0.25 ppm vs 0.36 ppm) of the susceptibility values obtained from imaging with protocols with different acquisition parameters such as spatial resolution and echo times. This improvement is especially important given that previous studies\cite{ZhouLiu2014, Karsa2019, ZhouColgan2019} suggested resolution-dependent bias.

\section*{Discussion and Conclusions}
In this work, we have described a new method for quantitative susceptibility mapping that leads to improved quantification of susceptibility in the liver for the assessment of iron overload. The proposed data adaptive method capitalizes on the fact that CSE imaging used to perform field mapping in the liver provides additional information that can be exploited to regularize the ill-conditioned dipole inversion . In particular, we propose two complementary ways to minimize sensitivity of the inversion to errors in the estimated field map and reduce artifacts in the susceptibility maps. The first feature of the proposed method is the use of the weighting map based on the size of the residual of CSE model fit, which reflects certainty of field map determination. The second feature is the use of FF map for segmentation of adipose tissue as zero-reference tissue to provide additional regularization of the dipole inversion problem and reduce shading artifact. We demonstrated that the proposed method shows higher correlations with two alternative measures of liver iron concentration, $R_2^*$ and $R_2$-based LIC, than the previously proposed method of Sharma et al\cite{Sharma2015}, which relies only on spatially constraining the dipole inversion problem. Additionally, the proposed data adaptive method exhibits high repeatability between test/re-test acquisitions of the same subject and reproducibility between two acquisition protocols with different parameters, including spatial resolution and echo times. The improvement of reproducibility is important as previous studies suggested a resolution-dependent bias in susceptibility measurements.
	
These performance improvements may be explained by the fact that the proposed adaptive weighting makes the reconstruction less sensitive to errors in estimated field maps, which are especially likely in the cases of high liver iron. At the same time, the use of adipose tissue as a zero reference helps reduce shading artifact, which improves repeatability and reproducibility. Since QSM can only determine susceptibility values relative to a reference, the use of adipose tissue as zero-reference has an added benefit of automatically performing the referencing step in QSM processing. Importantly, the proposed method does not require any additional acquisitions, as it utilizes information about FF and residual of CSE fit already available from CSE imaging for estimation of field map. Further, these auxiliary sources of information are inherently co-localized with the measured field map, as they are obtained from the same acquisition. 

	High linear correlation was observed between the proposed data adaptive QSM method and both $R_2^*$ ($r^2 = 0.98 -  0.99$) and $R_2$-based ($r^2 = 0.72 - 0.73$) measures of LIC. Lower correlation with $R_2$-based LIC may be explained by the fact that it provides a single global value of liver LIC, disregarding the potential heterogeneity of iron distribution in the liver, while $R_2^*$ measurements were performed in the same liver ROIs that were used for susceptibility quantification. Despite high correlation of susceptibility with $R_2^*$ measurements in the liver, there is evidence that QSM may provide a better measure of LIC as it is less sensitive than relaxometry to other issues such as microscopic distribution of iron\cite{Bashir2019, Colgan2020}. This hypothesis requires further evaluation, which, in turn, requires a robust, repeatable, and reproducible QSM algorithm.

	This work has several limitations. One limitation of the proposed method is that in the cases of extreme iron overload in the liver, the estimation of both field map and FF are challenging due to very fast $R_2^*$-induced signal decay even in the first echo, therefore these maps are likely to be unreliable and contain a large number of water/fat swaps. A possible solution would be to use an acquisition with shorter, closely spaced echo times to reduce the effects of $R_2^*$-induced decay. Another challenge may arise when using the proposed method in subjects with very little adipose tissue. In these cases, an alternative choice of zero-reference tissue may be needed. The study itself is also limited by the fact that all subjects were scanned at a single site on MRI scanners from a single vendor. Future work will assess repeatability and reproducibility of QSM methods using examinations performed at several sites. Finally, future studies would include measurements from biomagnetic liver susceptometry obtained with a superconducting quantum interference device (SQUID), which is considered the non-invasive “gold standard” to measure liver iron concentration but has extremely limited availability. 

	In conclusion, this work represents the next step towards establishing regularized abdominal QSM as an accurate, repeatable and reproducible technique for assessment of LIC in clinical settings. Our results indicate that data adaptive regularization incorporating data quality metrics and anatomical priors is a preferred approach for abdominal QSM as it (1) shows higher correlation with reference LIC values; (2) features an increase in repeatability and reproducibility of liver susceptibility measurements across two protocols with different acquisition parameters at 3.0T. 

\section*{Acknowledgements}
This work was supported by NIH (R01 DK117354, R01 DK100651, R01 DK088925, R01EB027087) and GE Healthcare.

\pagebreak

\end{document}